# Influence of the magnetron power on the Er-related photoluminescence of AlN:Er films prepared by magnetron sputtering

S.S. Hussain[1], V. Brien[1], H. Rinnert[2], P. Pigeat[1]

1. Institut Jean Lamour (UMR CNRS 7198), Département CP2S, CNRS – Nancy, Université - UPV-Metz, Boulevard des Aiguillettes B.P. 239, F-54506 Vandœuvre lès Nancy, France

2. Institut Jean Lamour (UMR CNRS 7198), Département P2M, CNRS – Nancy, Université - UPV-Metz, Boulevard des Aiguillettes B.P. 239, F-54506 Vandœuvre lès Nancy, France

The effect of magnetron power on the room temperature 1.54 μm infra-red photoluminescence intensity of erbium doped AlN films grown by r. f. magnetron sputtering, has been studied. The AlN:Er thin films were deposited on (001) Silicon substrates. The study presents relative photoluminescence intensities of nanocrystallized samples prepared with identical sputtering parameters for two erbium doping levels (0.5 and 1.5 atomic %). The structural evolution of the crystallites as a function of the power is followed by transmission electron microscopy.

**1 Introduction** For some time now, rare-earth (RE)-doped semiconductors represent significant potential applications in the field of opto-electronic technology. Part of this technological interest relies on the shielded 4f levels of the RE ions as they give rise to sharp and strong luminescence peaks [1-5]. Among the RE elements, Er is preferred to its counterparts since the Er ions can produce both visible light at 558 nm (green, one of the primary colours) and IR light at 1.54 μm whose spectrum region coincides with the main low-loss region in the absorption spectrum of silica-based optical fibres, combining so potential applications towards photonic devices and towards optical communication devices operating in the infrared domain. These interesting emissions can however only be exploited when placed into host matrixes. On one side, the shielding of the intra 4f levels prevents the shifting of the $RE^{3+}$ energy levels and ensures the frequency emission stability. Moreover the intra 4f transitions are parity forbidden for the isolated ions. Matrixes can render the $Er^{3+}$ ions optically active, via a relaxation of selection rules due to crystal field effects. As silicon based materials were tested in the 1960s to the 90s with no clear industrial success it was found that the thermal quenching effect in RE-doped semiconductors is related to the value of the optical band gap of the material [6]. Since then and within the scope of temperature-insensitive device application, $RE^{3+}$ ions have been extensively investigated in semiconductors and more especially in wider band gap compounds exhibiting lower temperature quenching: nitrided or oxided silicon compounds ($SiO_n$, SiN, $Si_3N_4$), ZnTe, GaP, or other nitrides: GeN, GaN, AlN…[7-15].

Amongst the III-V semiconductors exhibiting a direct band gap, AlN has received less attention than GaN. It however presents outstanding properties (larger gap of 6.2eV for the single crystal), excellent hardness, high thermal conductivity, high melting temperature and chemical stability... making it very promising for long-service life and reliable application. It has been shown that AlN:Er can exhibit significant photoluminescence at room temperature [12,15]. Although the physical mechanisms behind the excitation of the ions and the optical emission processes are not fully understood, the site in which erbium is located and its close atomic neighbours seems to be the key to control the PL efficiency. And these parameters can evolve with the morphology of the films. In of different erbium doped AlN matrixes prepared by magnetron sputtering with different morphologies.

The crystalline structure is observed by transmission electron microscopy (TEM). Compositions were obtained by electron dispersive spectroscopy of X-rays (EDSX) and Rutherford backscattering spectrometry (RBS). The photoluminescence (PL) emission in the visible and near infrared range was observed and recorded on the different AlN:Er samples and will be presented as a function of the magnetron power used for their synthesis.

## 2 Sample preparation and chemical analysis

AlN:Er films were deposited using a r. f. magnetron sputtering system at room temperature on Si (001) substrates in a gas mixture of Ar and $N_2$ of high purity (99.999 %). The $N_2/(Ar+N_2)$ percentage in the gas mixture was set to 50 %. The respective flows were set to 2.5 sccm $N_2$ and 2.5 sccm Ar in order to reach a total sputtering pressure P constant equal to 0.5 Pa. The system base pressure was equal to $1 \times 10^{-6}$ Torr. The insertion of Er (purity of 99.9 %) in samples was obtained by co-deposition by using a sectored aluminium (purity of 99.99 %) disk with adequate relative surfaces. Two targets were used to generate the 0.5 % and the 1.5 % erbium doped AlN samples. The diameter and the thickness of the targets were 60 mm and 3 mm large respectively and the target-sample distance was 150 mm. In this study, bias voltage was set to 0 Volt. The target was sputter cleaned for 15 min using an Ar plasma (cleaning conditions: W = 300 W, P = 0.5 Pa). The reactor was equipped with an interferential optical reflectometer for real-time control of the thickness and the growth rate of the deposited layer. A batch of six samples of identical thickness was prepared. The six depositions were made with identical process parameters: plasma working pressure = 0.5 Pa, 0 V substrate bias and different magnetron power W. The 0.5 % erbium and the 1.5 % erbium samples were prepared by using 50 W, 125 W, 300 W and 50 W, 200 W, 300 W, respectively.

EDSX analysis was used to get the erbium doping levels obtained in the samples. Although proper calibration of this technique was done (by using the stoichiometric compound $Er_2O_3$ as a reference to asses the Cliff Lorimer coefficients), the obtained results were independently checked by RBS. More details on the chemical analyses procedure can be found in Brien *et al*. [16]. EDSX analysis was performed by using an EDAX spectrometer mounted on a CM20 Philips microscope and equipped with an ultra thin window X-Ray detector. The analyses were carried out in nanoprobe mode with a diameter of the probe of 10 nm. Values of 0.5 atomic % (+/- 0.2%) and 1.5 % (+/- 0.2%) of erbium were obtained.

## 3 Results and discussion

**3.1 Structural characterisation of the films** The power is known to modify the structure of the AlN magnetron sputtered films [17, 18]. Considering that the low doping effect of erbium will have no visible effect on the growth mechanism of the films, the previous published works [17, 18] allow expecting that the samples prepared at high power are nano-columnar and that the ones prepared at low powers are nano-granular. Transmission Electron Microscopy (TEM) observations were performed on a PHILIPS CM20 microscope operating at an accelerating voltage of 200 kV by using the technique of microclivage of samples. Dark field images indicate that the films prepared by using a magnetron power of 50 W are made of nearly equiaxed (slightly elongated in the growth direction) nanograins (microscopy images chosen and shown in Fig. 1 correspond to cross sections views). The size of the grains lies in the 3 - 20 nm range. The images recorded on the films (Fig. 1.b and c) show that the AlN:Er films (regardless of their erbium content) synthesised by using a magnetron power W laying between 125 and 300 W are made of nano columns. It is rather difficult to discern an obvious evolution between the different microstructures, but the average width of the columns (laying in the range 15 – 30 nm) is actually a progressive and increasing function of the magnetron power.

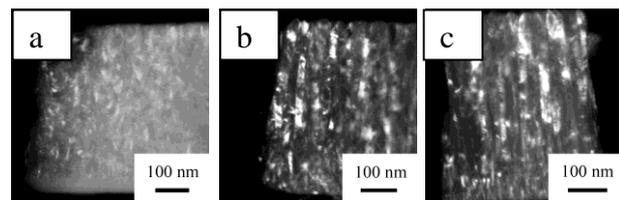

**Figure 1** Typical morphology of the films obtained on cross-viewed samples by TEM (micrographs are dark field images) (a) Samples prepared at 50 W, (b) Samples prepared at 125 W and 200 W, (c) Samples prepared at 300 W.

X-ray diffraction patterns were also recorded on the films but no direct relationship could be withdrawn between the power and the full width at half maximum of the different peaks, probably because of the presence of stresses in films. Full width at half maxima (FWHM) are not only related to the average width of the grains in the films but also to the state of stresses. Besides, when the power is modified, the energy given to the building atoms is increased, favouring the growth of textured films, which makes the interpretation of the FWHM X-ray diffraction peaks more complex.

**3.2 Photoluminescence results** The PL of the six AlN:Er films of this study was characterized in the range 450-1650 nm. PL experiments were performed at room temperature. For the steady state experiments in the visible range, the excitation was obtained with a 200 W mercury arc lamp source, using the ultraviolet (UV) lines at 313 and 334 nm. The PL signal was analyzed by a monochromator equipped with a 150 grooves / mm grating and by a charge-coupled device camera detector cooled at 140 K. For the experiments in the near-infrared domain, the sample was excited by a 30 mW He-Cd laser using the 325 nm line. The PL signal was analyzed by a monochromator equipped with a 600 grooves / mm grating and by a photomultiplier tube cooled at 190 K. The responses of the detection systems were precisely calibrated with a tungsten wire calibration source. The PL spectra of the samples doped with 1.5 Er at. % are given in Fig. 3 (a) and Fig. 3 (b) in the visible and near infrared range, respectively.

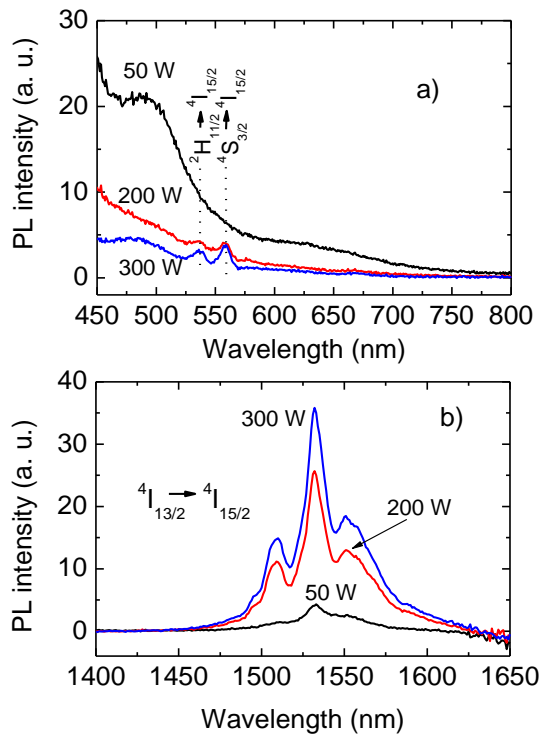

**Figure 3** Photoluminescence spectra of the Er-doped AlN films in the visible (a) and in the near infrared range (b). The samples contain around 1.5 at. % erbium.

The spectra show different contributions. A first wide band is obtained around 490 nm. The intensity of this PL band is a decreasing function of the magnetron power. This band is probably due to transitions between electronic states located in the band gap. It is well known that an amorphous or disordered structure involves in band tails states within semiconductors band gap. This band is not well defined and the maximum of this band seems to be at wavelengths lower than 450 nm, suggesting that the PL measured is the low energy contribution of this defect related PL band. Moreover, as the thickness of the layers is around 500 nm, an interference phenomenon can contribute to change the shape of this band from one sample to another [19]. Such a band has already been observed in many studies concerning the photoluminescence of AlN of different types: nano powders, nanostructured films… Although there is no consensus on the interpretation: only a few possibilities are mentioned: the PL large band is either said to be the result of native defects: deep-levels nitrogen vacancies, oxygen point defects substituting nitrogen atoms or the result of "surface-luminescence" of AlN nano particles [20-22]. In our case, the decrease of the PL intensity of this band as the magnetron power increases could be interpreted by the increase of the crystallized volume, as shown by the structural study. The oxygen content being equivalent in all samples, it cannot be considered here as a possible explanation. With increasing power, the grain boundaries volume fraction decreases, suggesting a decrease of disorder and a decrease of the number of electronic band tail states. This study would then prove that these bands are more related to band-edge or surface AlN particles (or grain boundaries) levels than to point defects levels.

Narrow PL bands are observed at 538, 558 and 1535 nm. These bands, which appear only in Er-doped samples, are attributed to the $^2H_{11/2} \rightarrow ^4I_{15/2}$, $^4S_{3/2} \rightarrow ^4I_{15/2}$ and $^4I_{13/2} \rightarrow ^4I_{15/2}$ intra 4f transitions. Similar PL band have been reported in Er doped c-axis oriented AlN films. The different narrow contributions observed, in particular for the near infrared PL band, are due to the crystal field splitting of the multiplets. These PL bands are strongly dependent on the magnetron power, as already observed by Liu et al. [23] for Tb doped AlN samples. For the sample prepared with a magnetron power equal to 50 W, the Er related visible PL band is not detectable and the 1.5 μm band is very weak. When the power is increased the green PL is clearly visible and the near infrared band is improved by almost one order of magnitude for the sample prepared with a power equal to 300 W. The Er-related PL intensity is then an increasing function of the power. This evolution has been confirmed with another Er concentration. The evolution of the integrated 1.5 μm PL intensity measured on the films containing two different Er contents is reported in Fig. 4 as a function of the power. Results have been weight by the coefficient of 0.5 or 1.5 corresponding to the erbium content so that the results can be read independently of the erbium concentration. As can be seen on the graph, the $Er^{3+}$ emission at 1.5 μm increases with the magnetron power used to prepare the samples, irrespectively of the erbium content (0.5 and 1.5 % erbium). Moreover, the two curves scale with the power in the same way: PL signals increase by the same rough factor of 8 from the one prepared at 50 W to the one prepared at 300 W.

As some authors evidenced that the presence of electronegative atoms, like oxygen, within erbium doped (gallium or aluminium) nitrides have a tangible effect on the efficiency of the ions luminescence [11, 24]. The oxygen content was measured in the six samples and was found to be the same amount around 10 atomic % in all samples.

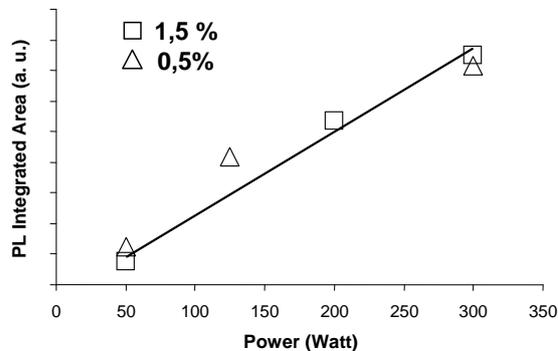

**Figure 4** 1.54 µm integrated PL intensity as a function of the magnetron power used to synthesise the AlN:Er films for the two chosen levels of erbium contents.

Consequently the oxygen may has an influence on the Er related PL but can not explain the observed dependence with the magnetron power. This implies that the only variable parameter from one sample to another is the power used to prepare the sample, or as shown thanks to the TEM investigation: their morphology. As the size of the AlN crystalline domain is also an increasing function of the magnetron power, and assuming the $Er^{3+}$ ions are randomly located, our result strongly suggests that a crystalline structure is favourable to the $Er^{3+}$ luminescence. It is known that the intra 4f transition of $Er^{3+}$ ions is parity forbidden. However, when the ions are located in particular lattice site having local symmetry, like tetrahedral or hexagonal symmetry, some electric dipole transitions can become allowed. The results could be interpreted by the existence of at least two different $Er^{3+}$ sites: one in the AlN crystalline column, the other one at the grain boundaries or in the disordered area. Ions located in crystalline environments would be preferentially optically active. However, the existence and influence of non-radiative centres on the PL cannot be ruled out. Further investigations are needed to investigate the possible effect of such defects.